\documentclass[twocolumn]{aa}
\usepackage{graphicx,natbib}
\usepackage{txfonts}
\bibpunct{(}{)}{;}{a}{}{,}
\begin{document}
   \title{Ruthenium and hafnium abundances \\
   in giant and dwarf barium stars
   \thanks{Based on spectroscopic observations collected at the
           European Southern Observatory (ESO), within the
           Observat\'orio Nacional ON/ESO and ON/IAG agreements,
           under FAPESP project n$^{\circ}$ 1998/10138-8.}}

%   \subtitle{}

   \author{D.M. Allen
          \inst{}\thanks{Current address: Centre for Astrophysics
      Research, STRI and School of Physics, Astronomy and
      Mathematics, University of Hertfordshire, Hatfield, UK.}
          \and
          G. F. Porto de Mello
      \inst{}
          }

   \offprints{D.M. Allen}

   \institute{Observat\'orio do Valongo/UFRJ,
              Ladeira do Pedro Antonio 43, 20080-090 Rio de Janeiro, RJ, Brazil
          \email{d.moreira-allen@herts.ac.uk, gustavo@ov.ufrj.br}
              }

   \date{Received, 2006; accepted, 2007}

% \abstract{}{}{}{}{}
% 5 {} token are mandatory

  \abstract
   {}
   {We present abundances for Ru and Hf,
   compare them to abundances of other heavy elements, and discuss the
   problems found in determining Ru and Hf abundances with laboratory
   $gf$-values in the spectra of barium stars.}
   {We determined Ru and Hf abundances in a sample
of giant and dwarf barium stars, by the spectral synthesis of two
\ion{Ru}{i} ($\lambda$4080.574 and $\lambda$4757.856) and two
\ion{Hf}{ii} ($\lambda$4080.437 and
$\lambda$4093.155) transitions. The stellar spectra were observed with
FEROS/ESO, and the stellar atmospheric parameters lie in the range
4300 $<$ $T_{\rm eff}$/K $<$ 6500, $-$1.2 $<$ [Fe/H] $\leq$ 0 and 1.4
$\leq \log g <$ 4.6.}
   {The \ion{Hf}{ii} $\lambda$4080 and the \ion{Ru}{i} $\lambda$4758 observed
transitions result in a unreasonably high solar abundance, 
given certain known uncertainties, when fitted with 
laboratory $gf$-values. For these two transitions we determined
empirical $gf$-values by fitting the observed line profiles of the
spectra of the Sun and Arcturus. For the sample stars, this procedure 
resulted in a good agreement of Ru and Hf
abundances given by the two available lines. The resulting Ru and
Hf abundances were compared to those of Y, Nd, Sm and Eu.
In the solar system Ru, Sm and Eu are dominated by the $r$-process 
and Hf, Nd and Y by the $s$-process, and  
all of these elements are enhanced in barium stars since they
lie inside the $s$-process path. 
Ru abundances show large scatter when compared to other
heavy elements, whereas Hf abundances show less scatter and
closely follow the abundances of Sm and Nd, in good agreement with
theoretical expectations. We also suggest a possible, unexpected,
correlation of Ru and Sm abundances. The observed behaviour in
abundances is probably due to variations in the $^{13}$C pocket
efficiency in AGB stars, and, though masked by high uncertainties,
hint at a more complex scenario than proposed by theory.}

  % conclusions heading (optional), 
   {}


%   \titilerunning
%   \authorrunning
   \maketitle
%
%________________________________________________________________

\section{Introduction}\label{int}

Barium stars are chemically peculiar objects which present large
excesses of the elements due to the neutron capture $s$-process.
These stars are not evolved enough to self-enrich during the
thermal pulses in the AGB phase. The standard explanation
for their peculiarities is a binary status. The former primary,
more massive, evolves faster and goes into the AGB phase, whereby
it convectively enriches its atmosphere with $s$-process products.
After a phase of strong mass loss through stellar wind, it becomes a
white dwarf and is detected in the ultraviolet only with
difficulty \citep{bohm2000}, if at all. The former secondary 
is enriched by mass accretion from the stellar wind of its
companion, and presents in its atmosphere vestiges of the
nucleosynthesis of the former AGB star, being presently observed as the
barium star.

Such stars usually have been studied by their strong
excesses of the heavy elements chiefly synthesized by the
$s$-process, Sr, Y, Zr, Ba, La, Ce and Nd being the most
spectroscopically accessible. Ruthenium and hafnium, in their
turn, have been very little studied even for this chemically
peculiar class of stars. In fact, the literature on
abundances of Ru and Hf, for any class of star, is very scarce.

These elements were previously detected in the very metal-poor
stars CS 22892-052 \citep[][and references therein]{sneden03} and
CS 31082-001 \citep{hill02}, thanks to their large enhancement of
neutron-capture elements. The lines of Hf and Ru are weak and most of
them lie
blended in the crowded near-UV region of the spectrum, generally
hampering their detection in normal stars.
\citeauthor{sneden03} found [Ru/Fe] = +1.34 and [Hf/Fe] = +1.24
for CS 22892-052, at [Fe/H] = $-$3.1 and \citeauthor{hill02} found
[Ru/Fe] = +1.42 and [Hf/Fe] = +1.43  for CS 31082-001, at [Fe/H] =
$-$2.9. Both these works only employed spectral lines with $\lambda$
$<$ 4000 $\rm \AA$. For another very metal poor star, HD 122563
with [Fe/H] = $-$2.7, \citet{honda06} found [Ru/Fe] = 0.07 from two
lines with $\lambda <$ 4000 $\rm \AA$. 
For the very metal poor ([Fe/H] = $-$3.5) TP-AGB star, 
CS 30322-023, \citet{mass06} found [Ru/Fe] = 1.05 and [Hf/Fe] = 0.53.
Ruthenium and hafnium were
also observed in some chemically peculiar stars. \citet{tl83}
found [Fe/H]=$-$0.32, [Ru/Fe]=+0.78 and [Hf/Fe]=+1.09 for the barium
star HR 774, from the $\lambda$4584 and $\lambda$5309 \ion{Ru}{i}
lines, and the $\lambda$7132 \ion{Hf}{ii} line. For the symbiotic
star AG Draconis, \citet{scjb96} found [Hf/Fe]=+0.86 and
[M/Fe]=$-$1.5 by using the $\lambda$7132 line. \citet{yush02} found
for the barium star $\zeta$ Cyg [Ru/Fe] $\approx$ $-$0.04 and
[Hf/Fe] $\approx$ 0.47, for which they also found
[\ion{Fe}{i}/H]=+0.02$\pm$0.10 and
[\ion{Fe}{ii}/H]=+0.06$\pm$0.08, employing the $\lambda$4584,
$\lambda$4869 and $\lambda$5026 \ion{Ru}{i} lines, and the
$\lambda$4093 \ion{Hf}{ii} line. We note that the lines of
\ion{Ru}{i} used by \citeauthor{yush02} are not visible in our
spectra.

Ruthenium lies close to Nb and Mo in the periodic table. This 
element could be called an $r$-process element, since,
according to \citet{arlandini99}, the contributions from the $s$-, 
$r$-, and $p$-processes for its abundance are respectively, 32.3\%, 
59.7\%, and 8\%.  
Hafnium is a heavy $s$-process element, lying close to La
and Ba, and the abundance contributions from the $s$-, $r$-, 
and $p$-processes are, respectively, 55.5\%, 44.16\% and 0.34\%, 
according to \citet{arlandini99}.
Abundance determinations of Ru and Hf for a
statistically significant sample of barium stars might help
increase the number of elements studied in these objects, thereby
shedding further light on the detailed abundance distribution of
heavy elements in these stars. The abundance ratio of the heavy
group of the $s$-process elements, Ba to Nd, to the lighter group,
Sr to Zr, has been traditionally used as a neutron exposure
parameter \citep{LB91,north94,di06b,rod07}. These data are necessary 
to clarify the mechanism that accounts for the large
spread in abundance excesses of neutron capture elements between
the mild barium and barium stars. The notion that these
differences could be accounted for by differing metallicities,
which in its turn results in different neutron exposure levels due
to a higher ratio of neutron to seed nuclei density for lower
metallicity stars, is not corroborated by the data
\citep{boyarchuk02,rod07}. A possible correlation with orbital
parameters, in the sense that the mild barium stars have
smaller $s$-process excesses due to longer orbital periods, has been
claimed \citep{BV84} but does not seem to explain the observations
either \citep{J98}. This has led \citet{rod07} to suggest that a
possible explanation might lie in a different mass range of the
progenitors of mild barium and barium stars, or in different
mixing properties in the two classes of stars. In order to test
these and other possible scenarios, detailed abundance data are
mandatory.

In this work, we present Ru and Hf abundances for a large sample
of dwarf barium stars, mild barium stars and classical barium
stars, based on high resolution and high signal-to-noise spectra.
We also discuss discrepancies in the Ru and Hf oscillator
strengths available in the literature, and their effect on the
derived abundances. This paper is organized as follows. Section
\ref{param} briefly presents the data and the determination
of the stellar atmospheric parameters; Sect. \ref{loggf} describes
the choice of spectral lines and the evaluation of published
laboratory oscillator strengths; in Sect. \ref{uncert} the
uncertainty calculations are detailed;
in Sect. \ref{abund} the derivation of abundances is
described, and in Sect. \ref{concl} our conclusions are drawn.

%______________________________________________________________

\section{Data source and atmospheric parameters}\label{param}

All spectra for the sample stars were obtained with the 1.52m
telescope at ESO, La Silla, using the Fiber Fed Extended Range
Optical Spectrograph \citep[FEROS,][]{kaufer00}. FEROS spectra
have a constant resolving power of R = 48000 from 3600 $\rm \AA$
to 9200 $\rm \AA$. The target sample for the present study
includes 25 dwarf, subgiant and giant barium stars from \citet{di06a}:
the S/N of the spectra of this sample ranges from 100 to 250.
To this we added the sample described in detail by
\citet{rod07}, involving 8 mild and classical barium
stars, plus 6 normal giants, with spectral S/N ratio ranging from
500 to 600. The journal of observations and derivation of
atmospheric parameters have been described in detail by
these authors.

\citet{di06a} determined effective temperatures $T_{\rm eff}$ from
photometry, surface gravities $\log g$ from the stellar positions in
theoretical HR diagrams, and iron abundances from equivalent widths of
approximately 150 \ion{Fe}{i} lines and 30 \ion{Fe}{ii} lines.
\citet{rod07} determined $T_{\rm eff}$, $\log g$ and metallicities
from the simultaneous excitation and ionization equilibria of the
equivalent widths of an average number of 120 \ion{Fe}{i} and 12
\ion{Fe}{ii} lines. Surface gravities were also computed from the
stellar luminosities and theoretical HR diagrams, and a very good
agreement was found for the two sets of gravities.

The LTE abundance analysis and the spectrum synthesis calculations
for Ru and Hf were performed by employing the codes by \citet[][
and subsequent improvements in the past thirty years]{Spi67},
described in \citet{cay91} and \citet{barb03}. The adopted model
atmospheres (NMARCS) were computed with a version of the MARCS
code, initially developed by \citet{gben75} and subsequently
updated by \citep{plez92}, used here for stars with gravities
$\log g$ $<$ 3.3, and \citet{edv93}, here used for less evolved
stars with $\log g$ $\geq$ 3.3. Abundances for Y, Nd, Sm, and Eu
were taken from \citet{di06a} and for stars of \citet{rod07},
abundances based on spectrum synthesis calculations were performed
and will be detailed in a forthcoming paper.

%______________________________________________________________

\section{The oscillator strengths of ruthenium and hafnium}\label{loggf}

\begin{figure}[ht!]
\centerline{\includegraphics[totalheight=9.0cm]{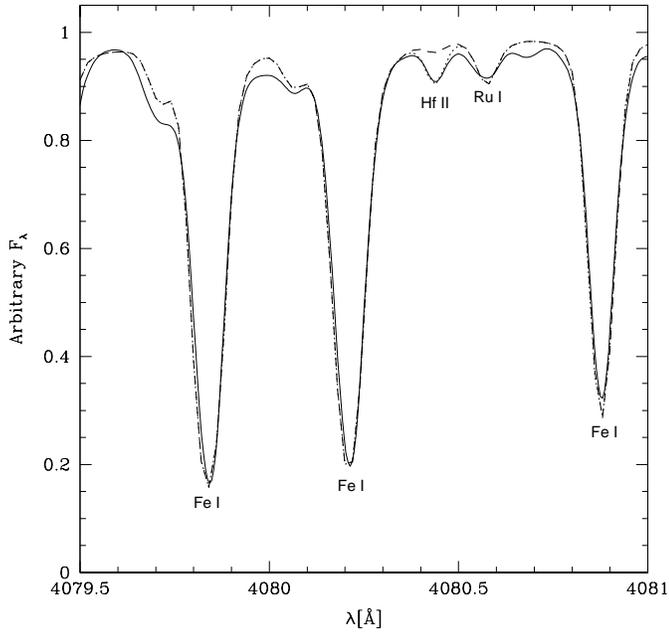}}
\caption{\label{6735fig1} Fits for two oscillator strengths
of the \ion{Hf}{ii} $\lambda$4080.437 line in the solar spectrum.
Solid line: observed spectrum; dotted line: $\log gf$ = $-$0.896;
dashed line: $\log gf$ = $-$1.596. All synthetic spectra were 
created with $\log\epsilon$(Hf) = 0.88.}
\end{figure}

\begin{figure}[ht!]
\centerline{\includegraphics[totalheight=9.0cm]{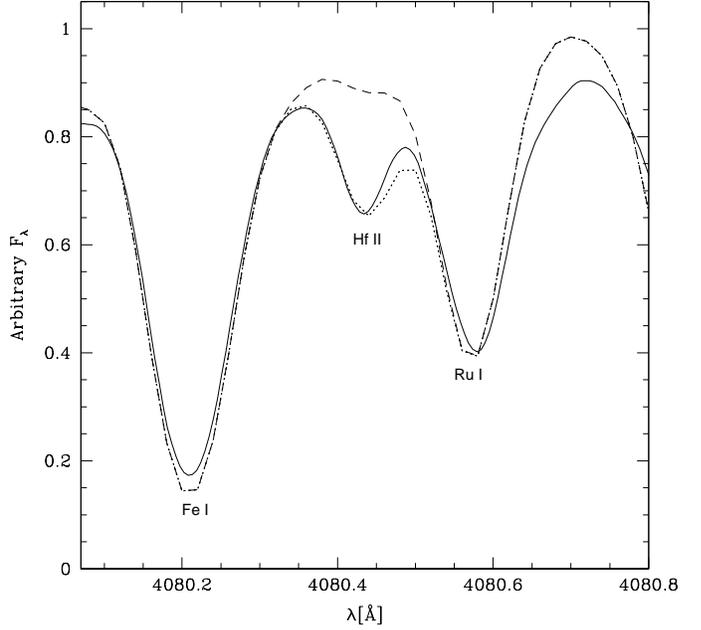}}
\caption{\label{6735fig2} Fits for two oscillator strengths
of the \ion{Hf}{ii} $\lambda$4080.437 line in the Arcturus
spectrum. Solid line: observed spectrum; dotted line:
$\log\epsilon$(Hf) = 0.28 ([Hf/Fe]=-0.06) and $\log gf$ = $-$0.896 or
$\log\epsilon$(Hf) = 0.98 ([Hf/Fe]=+0.64) and $\log gf$ = $-$1.596;
dashed line: $\log\epsilon$(Hf) = 0.28 and $\log gf$ = $-$1.596.}
\end{figure}

\begin{figure}[ht!]
\centerline{\includegraphics[totalheight=9.0cm]{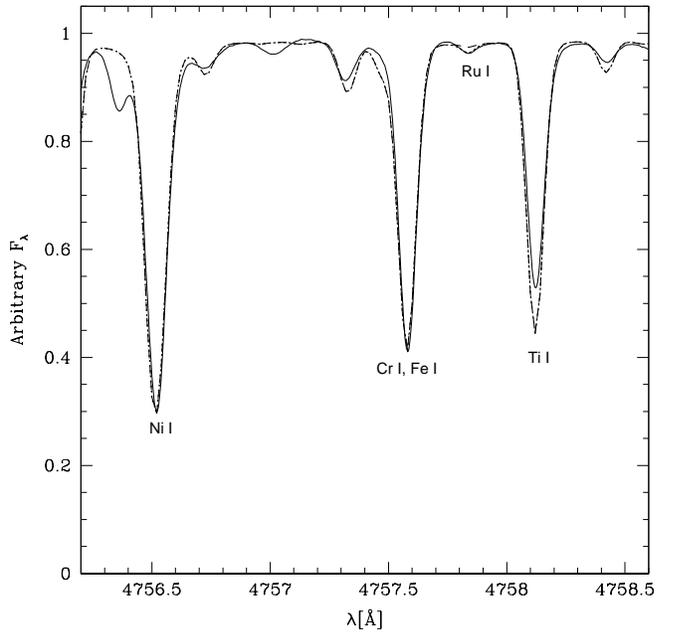}}
\caption{\label{6735fig3} Fits for two oscillator strengths
in the solar spectrum for the \ion{Ru}{i} $\lambda$4757.856 line.
Solid line: observed spectrum; dotted line: 
$\log\epsilon$(Ru) = 1.84 and $\log gf$ = $-$0.540 or 
$\log\epsilon$(Ru) = 2.19 and $\log gf$ = $-$0.890; 
dashed line: $\log\epsilon$(Ru) = 1.84 and $\log gf$ = $-$0.890.}
\end{figure}

Ruthenium and hafnium present spectral lines ranging from the UV
to the IR: for the lines that appear for $\lambda <$ 4000 $\rm
\AA$, the FEROS spectra do not allow good fits to synthetic ones,
this region being too crowded for cool stars. So, in this work we
looked for lines with larger wavelengths. The following lines,
detected in the spectra of the Sun \citep{kurucz84} and Arcturus
\citep{hinkle00}, were considered for abundance determinations in
the barium star spectra: $\lambda$4080.574, $\lambda$4144.1968,
$\lambda$4381.272, and $\lambda$4757.856 for \ion{Ru}{i} and
$\lambda$4080.437 and $\lambda$4093.155 for \ion{Hf}{ii}. The
lines $\lambda$4144.1968 and $\lambda$4381.272 are not visible in
our barium star spectra, and so were discarded. Table \ref{atcte}
shows the oscillator strengths ($\log gf$) and excitation
potential ($\chi_{ex}$) for the \ion{Ru}{i} and \ion{Hf}{ii} lines
used in this work, as well as the references for the $\log gf$
values. The main $\log gf$ source was the {\it Vienna Atomic Line
Database} \citep[VALD,][]{pisk95}. The $\log gf$ for \ion{Hf}{ii}
$\lambda$4080.437 line is given by \citet{lund06} and \citet{lawler07}. 
If the 
value of $-$1.596 is used, the solar abundance of Hf is 0.7 dex
higher than the value given by \citet{gs98}, as shown in Fig.
\ref{6735fig1}. For Arcturus (Fig. \ref{6735fig2}), the
resulting abundance is $\log\epsilon$(Hf) = 0.98 and
[Hf/Fe]=+0.64, much higher than expected for a normal, slightly
metal-poor giant like Arcturus. Similarly, if the VALD $\log gf$
value (-0.890) for the \ion{Ru}{i} $\lambda$4757.856 line is used,
the solar abundance of Ru is 0.35 dex higher than in
\citeauthor{gs98}, as shown in Fig. \ref{6735fig3}. These results
are deemed as unreasonable in the face of the known uncertainties.
For these two transitions we determined $gf$-values by fitting the
observed line profiles of the spectra of the Sun and Arcturus,
shown in Table \ref{atcte} and Figs. \ref{6735fig1},
\ref{6735fig2}, and \ref{6735fig3}. We used for the Sun
$T_{\rm eff}$ = 5780 K, $\log g$ = 4.44 (cgs) and a microturbulent
velocity of $\xi$ = 0.93 km/s, and an atmospheric model from
\citet{edv93}. For Arcturus, we used atmospheric parameters from
\citet{mel03}, $T_{\rm eff}$ = 4275 K, $\log g$ = 1.55 (cgs),
[Fe/H] = $-$0.54, excepting $\xi$ = 1.5 km/s which was
fitted in this work. The \citet{plez92} grid was used to
derive the atmospheric model.

It is conceivable that an unidentified transition, blended with
the Ru and Hf lines, explains these large discrepancies. The
region around $\lambda$4080 contains molecular lines of CN, so one
must be careful in fitting synthetic spectra there. One
possibility is that a CN feature, missing in our molecular lines
database, might be blended with the \ion{Hf}{ii} $\lambda$4080.574
transition. Using the $gf$-values given by \citet{lund06} or \citet{lawler07}
could make the synthetic line weaker 
than the observed one, consequently increasing the resulting 
abundance from this line.

%----------Table atcte-----------------------------------------------------------------------------
\begin{table}
\caption{Atomic constants for the \ion{Ru}{i} and \ion{Hf}{ii}
lines used in this work, and references for $\log gf$. ``Sun''
denotes the $gf$-values fitted on the solar and Arcturus spectra.
Other $gf$-values were taken from VALD, \citet{lund06} (L06),
and \citet{lawler07} (L07).}
\label{atcte}
\setlength\tabcolsep{5pt}
\begin{tabular}{lcccccc}
\hline\hline
\noalign{\smallskip}
 el &  $\lambda$ (\rm \AA) & $\chi_{ex} (eV)$ & $\log gf$ & $\log gf$ & $\log gf$ & $\log gf$ \\
    &            &           & VALD   &  Sun & L06 & L07\\
\hline
\noalign{\smallskip}
Hf II & 4080.437 & 0.608 & ...    & -0.896 & -1.596  & -1.55 \\
Hf II & 4093.155 & 0.452 & -1.090 & ...    & ...     & -1.15 \\
Ru I  & 4080.574 & 0.810 & -0.040 & ...    & ...     &  ...  \\
Ru I  & 4757.856 & 0.928 & -0.890 & -0.540 & ...     &  ...  \\
\noalign{\smallskip}
\hline
\end{tabular}
\end{table}
%---------------------------------------------------------------------------------------

%
%______________________________________________________________

%------------Table Uncertainties------------------------------------------------------------
\begin{table}
\caption{Uncertainties on abundances. $\log{A_{pf}}$: output with
the atmospheric parameters adopted; $\log{A_{pT}}$: output by
altering 1 $\sigma$ on adopted $T\rm _{eff}$; $\log{A_{pmt}}$:
output by altering 1 $\sigma$ on adopted metallicity;
$\log{A_{pl}}$: output by altering 1 $\sigma$ on adopted $\log g$;
$\log{A_{p\xi}}$: output by altering 1 $\sigma$ on adopted
microturbulent velocity $\xi$; $\sigma_l$:
$\sigma_{\log\epsilon}$, the uncertainty on $\log\epsilon$(X) from
Table \ref{abuncom}; $\sigma_f$: $\sigma_{[X/Fe]}$, the
uncertainty on [Ru/Fe] or [Hf/Fe] from Table \ref{abuncom}.}
\label{errab} \setlength\tabcolsep{1.5pt}
\begin{tabular}{ccccccccc}
\hline\hline
 el & $\lambda$ ($\rm \AA$) & $\log{A_{pf}}$ & $\log{A_{pT}}$ & $\log{A_{pmt}}$ & $\log{A_{pl}}$ &
$\log{A_{p\xi}}$ & $\sigma_l$ & $\sigma_f$ \\
\hline
 & & & HD181053 & & & \\
\hline
Ru & 4080.574 & 1.99 & 2.01 & 1.91 & 2.01 & 1.99 & 0.47 & 0.46 \\
Ru & 4757.856 & 2.09 & 2.19 & 1.99 & 2.11 & 2.09 & & \\
Hf & 4080.437 & 1.13 & 1.14 & 1.13 & 1.28 & 1.13 & 0.17 & 0.15 \\
Hf & 4093.155 & 1.18 & 1.19 & 1.18 & 1.33 & 1.18 & & \\
\noalign{\smallskip}
\hline\hline
 & & & HD 87080 & & & \\
\hline
Ru & 4080.574 & 3.24 & 3.34 & 3.20 & 3.24 & 3.19 & 0.09 & 0.11 \\
Ru & 4757.856 & 3.44 & 3.54 & 3.40 & 3.44 & 3.44 & & \\
Hf & 4080.437 & 2.68 & 2.70 & 2.64 & 2.72 & 2.63 & 0.06 & 0.09 \\
Hf & 4093.155 & 2.48 & 2.50 & 2.44 & 2.52 & 2.46 & & \\
\noalign{\smallskip}
\hline
\end{tabular}
\\
\end{table}
%-------------------------------------------------------------------------------

\section{Uncertainties}\label{uncert}

Two stars were used to compute the abundance uncertainties: HD
181053 from \citet{rod07} with $T\rm _{eff}$ = 4810 $\pm$ 50 K,
$\log g$ = 2.48 $\pm$ 0.35, [Fe/H] = $-$0.19 $\pm$ 0.12, and $\xi$ =
1.70 $\pm$ 0.06, and HD 87080 from \citet{di06a} with
$T\rm _{eff}$ = 5460 $\pm$ 100 K, $\log g$ = 3.7 $\pm$ 0.2, [Fe/H] = $-$0.44
$\pm$ 0.04, and $\xi$ = 1.0 $\pm$ 0.1. The abundance uncertainties
were calculated by verifying how much the variation of 1 $\sigma$
on the atmospheric parameters affects the output value of the
synthesis program, here $\log{A_p}$. Table
\ref{errab} shows the values taken into account in this
calculation and the resulting uncertainties.

Under the simplifying hypothesis of independent errors, the
uncertainty of the output value is given by

\begin{equation}
\label{erapinst}
\sigma_{Ap}=\sqrt{(\Delta A_T)^2+(\Delta A_{mt})^2+(\Delta A_l)^2+(\Delta A_\xi)^2},
\end{equation}
where $\Delta A_T$, $\Delta A_{mt}$, $\Delta A_l$, and $\Delta A_\xi$, are the average
differences in $A_p$ from two lines of each element shown in Table \ref{errab}
due to variations of 1$\sigma$ in the temperature, metallicity, $\log g$, 
and microturbulent velocity, respectively.

The average value of $A_p$ ($A_{pm}$) is obtained by averaging the
individual abundances of two lines and not from several
measurements of the same line. In the latter case, the standard
deviation could be used to calculate the uncertainty on $A_{pm}$.
Considering this, we found it more suitable to apply a
propagation of errors taking into account the uncertainty
calculated with Eq. \ref{erapinst}. Thus, the uncertainty on
$A_{pm}$ is
\begin{equation}
\sigma_{Apm}={\sigma_{Ap}\over \sqrt{n}},
\end{equation}
where $n$ is the number of lines. The uncertainty on the logarithm of $A_{pm}$ is
\begin{equation}
\sigma_{\log(Apm)}={\sigma_{Apm}\over A_{pm}\ln{10}}.
\end{equation}

The abundance $\log\epsilon$(X) is related to the output of the synthesis program by
$\log\epsilon$(X) = $\log{A_{pm}}$ + [Fe/H]. Therefore, the uncertainty is

\begin{equation}
\sigma_{\log\epsilon(X)}=\sqrt{\sigma_{\log{(Apm)}}^2+\sigma_{\rm [Fe/H]}^2}.
\end{equation}

The relation between the abundance excess relative to iron [X/Fe] and the
output value of the synthesis program is
[X/Fe] = $\log{A_{pm}}$ - $\log\epsilon_\odot(X)$, where $\log\epsilon_\odot(X)$
is the solar abundance of the element ``X''. The uncertainty is calculated by

\begin{equation}
\sigma_{\rm [X/Fe]}=\sqrt{\sigma_{\log(Apm)}^2+\sigma_{\log\epsilon_\odot(X)}^2}.
\end{equation}

The uncertainties shown in Table \ref{errab} for HD 181053 and HD
87080 are typical for stars with $\log g <$ 3.3 and
$\log g \geq$ 3.3, respectively. Figures \ref{6735fig4} and
\ref{6735fig5} show the synthetic spectra considering 1 $\sigma$
of the abundance and Figs. \ref{6735fig6} and \ref{6735fig7} show
the maximum value of uncertainties on each axis.

For [Hf/Ru] the uncertainties are determined by
\begin{equation}
\sigma_{\rm [Hf/Ru]}=\sqrt{\sigma_{\rm [Hf/Fe]}^2+\sigma_{\rm [Ru/Fe]}^2},
\end{equation}
and they are 0.48 and 0.14 for stars with $\log g <$ 3.3 and
$\log g \geq$ 3.3, respectively.

Uncertainties on elements other than Ru and Hf were described in
\citet{di06a}.

%
%______________________________________________________________

%------------Table eqwid------------------------------------------------------------
\begin{table*}
\caption{Equivalent Width and abundance results for the sample stars, line by line.
The symbol '$<$' indicates an upper limit. The stars signaled with '*' were
considered normal rather than barium stars by \citet{rod07}.}
{\scriptsize \label{eqwid}
   $$
\setlength\tabcolsep{3pt}
\begin{tabular}{llrrrrrcllrrrrr}
\hline\hline
\noalign{\smallskip}
\multicolumn{2}{c}{identifiers} & el & $\lambda$ (\rm \AA) & EW & $\log\epsilon$(X) & [X/Fe] &&
\multicolumn{2}{c}{identifiers} & el & $\lambda$ (\rm \AA) & EW & $\log\epsilon$(X) & [X/Fe] \\
&&& (m\rm \AA) &&& &&&& (m\rm \AA) && \\
\noalign{\smallskip}
\hline
HD 749      &         & Ru I  & 4080.574 &  64 &     2.48 &    0.70 && HD 106191   &         & Hf II & 4080.437 & ... &     ...  &     ... \\
HD 749      &         & Ru I  & 4757.856 &  52 &     2.43 &    0.65 && HD 106191   &         & Hf II & 4093.155 & ... &  $<$1.09 & $<$0.50 \\
HD 749      &         & Hf II & 4080.437 &  56 &     2.27 &    1.45 && HD 107574   &         & Ru I  & 4080.574 & ... &      ... &     ... \\
HD 749      &         & Hf II & 4093.155 & 107 &     2.27 &    1.45 && HD 107574   &         & Ru I  & 4757.856 &  11 &     3.29 &    2.00 \\
HD 2454     & HR 107  & Ru I  & 4080.574 & ... &  $<$2.48 & $<$1.00 && HD 107574   &         & Hf II & 4080.437 & ... & ...      &    ...  \\
HD 2454     & HR 107  & Ru I  & 4757.856 &   3 &     2.88 &    1.40 && HD 107574   &         & Hf II & 4093.155 &  13 &     1.33 &    1.00 \\
HD 2454     & HR 107  & Hf II & 4080.437 & ... &  $<$0.92 & $<$0.40 && HD 113226*  & HR 4932 & Ru I  & 4080.574 &  31 &     1.86 &   -0.10 \\
HD 2454     & HR 107  & Hf II & 4093.155 & ... &  $<$0.52 & $<$0.00 && HD 113226*  & HR 4932 & Ru I  & 4757.856 &  12 &     1.96 &    0.00 \\
HD 5424     &         & Ru I  & 4080.574 &  71 &     2.19 &    0.90 && HD 113226*  & HR 4932 & Hf II & 4080.437 &   9 &     0.80 &   -0.20 \\
HD 5424     &         & Ru I  & 4757.856 &  77 &     2.49 &    1.20 && HD 113226*  & HR 4932 & Hf II & 4093.155 &  50 &     0.80 &   -0.20 \\
HD 5424     &         & Hf II & 4080.437 &  92 &     2.08 &    1.75 && HD 116713   & HR 5058 & Ru I  & 4080.574 & 105 &     2.42 &    0.70 \\
HD 5424     &         & Hf II & 4093.155 & ... &      ... &     ... && HD 116713   & HR 5058 & Ru I  & 4757.856 &  77 &     2.62 &    0.90 \\
HD 8270     & HR 391  & Ru I  & 4080.574 &   8 &     2.37 &    0.95 && HD 116713   & HR 5058 & Hf II & 4080.437 &  75 &     1.76 &    1.00 \\
HD 8270     & HR 391  & Ru I  & 4757.856 &   6 &     2.42 &    1.00 && HD 116713   & HR 5058 & Hf II & 4093.155 & 117 &     1.76 &    1.00 \\
HD 8270     & HR 391  & Hf II & 4080.437 &   9 &     1.16 &    0.70 && HD 116869   &         & Ru I  & 4080.574 &  59 &     2.02 &    0.50 \\
HD 8270     & HR 391  & Hf II & 4093.155 &  11 &     1.36 &    0.90 && HD 116869   &         & Ru I  & 4757.856 &  46 &     2.22 &    0.70 \\
HD 9362*    & HR 440  & Ru I  & 4080.574 &  25 &     1.55 &    0.05 && HD 116869   &         & Hf II & 4080.437 &  68 &     1.51 &    0.95 \\
HD 9362*    & HR 440  & Ru I  & 4757.856 &  15 &     1.75 &    0.25 && HD 116869   &         & Hf II & 4093.155 &  63 &     1.26 &    0.70 \\
HD 9362*    & HR 440  & Hf II & 4080.437 &  16 &     0.69 &    0.15 && HD 123396   &         & Ru I  & 4080.574 &  58 &     1.45 &    0.80 \\
HD 9362*    & HR 440  & Hf II & 4093.155 &  35 &     0.54 &    0.00 && HD 123396   &         & Ru I  & 4757.856 &  48 &     1.65 &    1.00 \\
HD 12392    &         & Ru I  & 4080.574 &  76 &     3.22 &    1.50 && HD 123396   &         & Hf II & 4080.437 &  72 &     1.39 &    1.70 \\
HD 12392    &         & Ru I  & 4757.856 &  61 &     3.02 &    1.30 && HD 123396   &         & Hf II & 4093.155 &  87 &     1.59 &    1.90 \\
HD 12392    &         & Hf II & 4080.437 &  63 &     2.42 &    1.66 && HD 139195   & HR 5802 & Ru I  & 4080.574 &  39 &     2.07 &    0.25 \\
HD 12392    &         & Hf II & 4093.155 &  95 &     2.42 &    1.66 && HD 139195   & HR 5802 & Ru I  & 4757.856 &  20 &     2.07 &    0.25 \\
HD 13551    &         & Ru I  & 4080.574 &   8 &     2.40 &    1.00 && HD 139195   & HR 5802 & Hf II & 4080.437 &  22 &     1.06 &    0.20 \\
HD 13551    &         & Ru I  & 4757.856 &   8 &     2.70 &    1.30 && HD 139195   & HR 5802 & Hf II & 4093.155 &  49 &     1.16 &    0.30 \\
HD 13551    &         & Hf II & 4080.437 &  21 &     1.39 &    0.95 && HD 147609   &         & Ru I  & 4080.574 & ... &  $<$2.69 & $<$1.30 \\
HD 13551    &         & Hf II & 4093.155 &  23 &     1.54 &    1.10 && HD 147609   &         & Ru I  & 4757.856 &  15 &     3.09 &    1.70 \\
HD 13611*   & HR 649  & Ru I  & 4080.574 &  23 &     1.95 &    0.25 && HD 147609   &         & Hf II & 4080.437 & ... &  $<$2.03 & $<$1.60 \\
HD 13611*   & HR 649  & Ru I  & 4757.856 &  19 &     2.25 &    0.55 && HD 147609   &         & Hf II & 4093.155 & ... &      ... &     ... \\
HD 13611*   & HR 649  & Hf II & 4080.437 &  33 &     0.99 &    0.25 && HD 150862   &         & Ru I  & 4080.574 & ... &      ... &     ... \\
HD 13611*   & HR 649  & Hf II & 4093.155 &  67 &     0.94 &    0.20 && HD 150862   &         & Ru I  & 4757.856 &   1 &     2.99 &    1.25 \\
HD 20894*   & HR 1016 & Ru I  & 4080.574 &  17 &     1.88 &    0.15 && HD 150862   &         & Hf II & 4080.437 & ... &      ... &     ... \\
HD 20894*   & HR 1016 & Ru I  & 4757.856 &  13 &     2.08 &    0.35 && HD 150862   &         & Hf II & 4093.155 &  12 &     1.48 &    0.70 \\
HD 20894*   & HR 1016 & Hf II & 4080.437 &  18 &     0.92 &    0.15 && HD 181053   & HR 5802 & Ru I  & 4080.574 &  47 &     1.80 &    0.15 \\
HD 20894*   & HR 1016 & Hf II & 4093.155 &  68 &     1.28 &    0.51 && HD 181053   & HR 5802 & Ru I  & 4757.856 &  27 &     1.90 &    0.25 \\
HD 22589    &         & Ru I  & 4080.574 &  12 &     2.12 &    0.55 && HD 181053   & HR 5802 & Hf II & 4080.437 &  30 &     0.94 &    0.25 \\
HD 22589    &         & Ru I  & 4757.856 &   5 &     2.27 &    0.70 && HD 181053   & HR 5802 & Hf II & 4093.155 &  71 &     0.99 &    0.30 \\
HD 22589    &         & Hf II & 4080.437 &  18 &     1.06 &    0.45 && HD 188985   &         & Ru I  & 4080.574 & ... &      ... &     ... \\
HD 22589    &         & Hf II & 4093.155 &  49 &     1.31 &    0.70 && HD 188985   &         & Ru I  & 4757.856 &  13 &     3.09 &    1.55 \\
HD 26967*   & HR 1326 & Ru I  & 4080.574 & 145 &     1.84 &    0.00 && HD 188985   &         & Hf II & 4080.437 & ... &      ... &     ... \\
HD 26967*   & HR 1326 & Ru I  & 4757.856 &  87 &     1.84 &    0.00 && HD 188985   &         & Hf II & 4093.155 &  34 &     1.68 &    1.10 \\
HD 26967*   & HR 1326 & Hf II & 4080.437 &  92 &     0.78 &   -0.10 && HD 202109   & HR 8115 & Ru I  & 4080.574 &  42 &     1.90 &    0.10 \\
HD 26967*   & HR 1326 & Hf II & 4093.155 &  90 &     0.78 &   -0.10 && HD 202109   & HR 8115 & Ru I  & 4757.856 &  26 &     2.05 &    0.25 \\
HD 27271    &         & Ru I  & 4080.574 &  58 &     2.20 &    0.45 && HD 202109   & HR 8115 & Hf II & 4080.437 &  27 &     0.99 &    0.15 \\
HD 27271    &         & Ru I  & 4757.856 &  29 &     2.05 &    0.30 && HD 202109   & HR 8115 & Hf II & 4093.155 &  93 &     1.09 &    0.25 \\
HD 27271    &         & Hf II & 4080.437 &  73 &     1.49 &    0.70 && HD 204075   & HR 8204 & Ru I  & 4080.574 & ... &      ... &     ... \\
HD 27271    &         & Hf II & 4093.155 &  67 &     1.49 &    0.70 && HD 204075   & HR 8204 & Ru I  & 4757.856 &  60 &     3.50 &    1.75 \\
HD 46407    & HR 2392 & Ru I  & 4080.574 &  87 &     2.70 &    0.95 && HD 204075   & HR 8204 & Hf II & 4080.437 & ... &      ... &     ... \\
HD 46407    & HR 2392 & Ru I  & 4757.856 &  74 &     2.90 &    1.15 && HD 204075   & HR 8204 & Hf II & 4093.155 & 118 &     1.79 &    1.00 \\
HD 46407    & HR 2392 & Hf II & 4080.437 &  86 &     1.94 &    1.15 && HD 205011   &         & Ru I  & 4080.574 &  66 &     2.05 &    0.35 \\
HD 46407    & HR 2392 & Hf II & 4093.155 &  92 &     1.94 &    1.15 && HD 205011   &         & Ru I  & 4757.856 &  40 &     2.15 &    0.45 \\
HD 48565    &         & Ru I  & 4080.574 &  12 &     2.52 &    1.30 && HD 205011   &         & Hf II & 4080.437 &  46 &     1.29 &    0.55 \\
HD 48565    &         & Ru I  & 4757.856 &  15 &     2.97 &    1.75 && HD 205011   &         & Hf II & 4093.155 &  92 &     1.34 &    0.60 \\
HD 48565    &         & Hf II & 4080.437 &  34 &     1.66 &    1.40 && HD 210709   &         & Ru I  & 4080.574 &  67 &     2.00 &    0.20 \\
HD 48565    &         & Hf II & 4093.155 &  25 &     1.66 &    1.40 && HD 210709   &         & Ru I  & 4757.856 &  40 &     2.30 &    0.50 \\
HD 76225    &         & Ru I  & 4080.574 &   3 &     2.83 &    1.30 && HD 210709   &         & Hf II & 4080.437 &  36 &     1.34 &    0.50 \\
HD 76225    &         & Ru I  & 4757.856 &  10 &     3.23 &    1.70 && HD 210709   &         & Hf II & 4093.155 &  64 &     1.04 &    0.20 \\
HD 76225    &         & Hf II & 4080.437 &  33 &     1.67 &    1.10 && HD 210910   &         & Ru I  & 4080.574 & ... &  $<$1.47 & $<$0.00 \\
HD 76225    &         & Hf II & 4093.155 &  20 &     1.47 &    0.90 && HD 210910   &         & Ru I  & 4757.856 & ... &     1.47 &    0.00 \\
HD 87080    &         & Ru I  & 4080.574 &  28 &     2.80 &    1.40 && HD 210910   &         & Hf II & 4080.437 & ... &  $<$0.81 & $<$0.30 \\
HD 87080    &         & Ru I  & 4757.856 &  29 &     3.00 &    1.60 && HD 210910   &         & Hf II & 4093.155 & ... &  $<$0.51 & $<$0.00 \\
HD 87080    &         & Hf II & 4080.437 &  51 &     2.24 &    1.80 && HD 220009*  & HR 8878 & Ru I  & 4080.574 &  53 &     1.27 &    0.10 \\
HD 87080    &         & Hf II & 4093.155 &  34 &     2.04 &    1.60 && HD 220009*  & HR 8878 & Ru I  & 4757.856 &  27 &     1.22 &    0.05 \\
HD 89948    &         & Ru I  & 4080.574 &   5 &     2.54 &    1.00 && HD 220009*  & HR 8878 & Hf II & 4080.437 &  21 &     0.21 &    0.00 \\
HD 89948    &         & Ru I  & 4757.856 &   8 &     2.79 &    1.25 && HD 220009*  & HR 8878 & Hf II & 4093.155 &  50 &     0.21 &    0.00 \\
HD 89948    &         & Hf II & 4080.437 &  17 &     1.48 &    0.90 && HD 222349   &         & Ru I  & 4080.574 &  16 &     2.71 &    1.50 \\
HD 89948    &         & Hf II & 4093.155 &  19 &     1.38 &    0.80 && HD 222349   &         & Ru I  & 4757.856 &   5 &     3.01 &    1.80 \\
HD 92545    &         & Ru I  & 4080.574 & ... &      ... &     ... && HD 222349   &         & Hf II & 4080.437 &  17 &     1.55 &    1.30 \\
HD 92545    &         & Ru I  & 4757.856 &   5 &     3.02 &    1.30 && HD 222349   &         & Hf II & 4093.155 &  21 &     1.55 &    1.30 \\
HD 92545    &         & Hf II & 4080.437 & ... &      ... &     ... && BD+18 5215  &         & Ru I  & 4080.574 &   3 &     2.81 &    1.50 \\
HD 92545    &         & Hf II & 4093.155 &  11 &     1.26 &    0.50 && BD+18 5215  &         & Ru I  & 4757.856 & ... &      ... &     ... \\
HD 104979   & HR 4608 & Ru I  & 4080.574 &  47 &     2.04 &    0.55 && BD+18 5215  &         & Hf II & 4080.437 &  20 &     1.55 &    1.20 \\
HD 104979   & HR 4608 & Ru I  & 4757.856 &  33 &     2.24 &    0.75 && BD+18 5215  &         & Hf II & 4093.155 &  15 &     1.55 &    1.20 \\
HD 104979   & HR 4608 & Hf II & 4080.437 &  54 &     1.18 &    0.65 && HD 223938   &         & Ru I  & 4080.574 &  38 &     1.99 &    0.50 \\
HD 104979   & HR 4608 & Hf II & 4093.155 &  77 &     1.08 &    0.55 && HD 223938   &         & Ru I  & 4757.856 &  23 &     2.24 &    0.75 \\
HD 106191   &         & Ru I  & 4080.574 & ... &      ... &     ... && HD 223938   &         & Hf II & 4080.437 &  36 &     1.58 &    1.05 \\
HD 106191   &         & Ru I  & 4757.856 &   8 &     3.25 &    1.70 && HD 223938   &         & Hf II & 4093.155 &  58 &     1.58 &    1.05 \\
\noalign{\smallskip}
\hline
\end{tabular}
   $$
}
$\log\epsilon$(X)=($\log{n_X/n_H}$)+12 and [X/Fe]=$\log\epsilon$(X)$_\ast$-$\log\epsilon$(X)$_\odot$-[Fe/H]
\end{table*}
%---------------------------------------------------------------------------------------

%------------Table abuncom------------------------------------------------------------
\begin{table*}
\caption{[Fe/H] and the mean values for $\log\epsilon$(X) and [X/Fe] for all stars of the
sample. The symbol '$<$' indicates an upper limit. The stars signaled with '*' were
considered normal rather than barium stars by \citet{rod07}. The number of lines used to 
compute the medium is shown in brackets.}
\label{abuncom}
   $$
\setlength\tabcolsep{3pt}
\begin{tabular}{llrrrrrrrrrr}
\hline\hline
\noalign{\smallskip}
\multicolumn{2}{c}{identifiers} & [Fe/H] & $\log\epsilon$(Ru) & $\log\epsilon$(Hf) & [Ru/Fe] & [Hf/Fe] &
[Y/Fe] & [Nd/Fe] & [Eu/Fe] & [Sm/Fe] & [Hf/Ru] \\
\noalign{\smallskip}
\hline
HD 749     &         & -0.06 & 2.46 &    2.27 &  0.68 &    1.45 &  1.18[12] &  1.33[9] &  0.33[4] &  0.93[5] &  0.77 \\
HD 2454    & HR 107  & -0.36 & 2.88 & $<$0.76 &  1.40 & $<$0.24 &  0.60[12] &  0.32[9] &  0.04[2] &  0.30[5] & -1.16 \\
HD 5424    &         & -0.55 & 2.36 &    2.08 &  1.07 &    1.75 &  1.03[11] &  1.72[9] &  0.46[4] &  1.27[4] &  0.68 \\
HD 8270    & HR 391  & -0.42 & 2.40 &    1.27 &  0.98 &    0.81 &  0.95[12] &  0.73[9] &  0.32[4] &  0.37[5] & -0.17 \\
HD 9362*   & HR 440  & -0.34 & 1.66 &    0.62 &  0.16 &    0.08 & -0.15[11] & -0.02[9] &  0.14[4] &  0.05[5] & -0.08 \\
HD 12392   &         & -0.12 & 3.13 &    2.42 &  1.41 &    1.66 &  1.21[12] &  1.49[9] &  0.48[4] &  1.47[5] &  0.25 \\
HD 13551   &         & -0.44 & 2.58 &    1.47 &  1.18 &    1.03 &  1.08[12] &  0.73[8] &  0.21[3] &  0.45[4] & -0.15 \\
HD 13611*  & HR 649  & -0.14 & 2.13 &    0.97 &  0.43 &    0.23 & -0.01[12] &  0.19[9] &  0.25[4] &  0.21[5] & -0.20 \\
HD 20894*  & HR 1016 & -0.11 & 1.99 &    1.14 &  0.26 &    0.37 & -0.04[12] &  0.10[9] &  0.18[4] &  0.08[5] &  0.11 \\
HD 22589   &         & -0.27 & 2.20 &    1.20 &  0.63 &    0.59 &  0.83[12] &  0.32[9] &  0.21[4] &  0.08[5] & -0.04 \\
HD 26967*  & HR 1326 &  0.00 & 1.84 &    0.78 &  0.00 &   -0.10 & -0.15[11] & -0.10[9] &  0.07[4] & -0.05[5] & -0.10 \\
HD 27271   &         & -0.09 & 2.13 &    1.49 &  0.38 &    0.70 &  0.89[12] &  0.53[9] &  0.31[4] &  0.41[5] &  0.32 \\
HD 46407   & HR 2392 & -0.09 & 2.81 &    1.94 &  1.06 &    1.15 &  1.15[12] &  0.88[9] &  0.34[4] &  0.89[5] &  0.09 \\
HD 48565   &         & -0.62 & 2.80 &    1.66 &  1.58 &    1.40 &  1.01[12] &  1.31[9] &  0.35[4] &  0.95[5] & -0.18 \\
HD 76225   &         & -0.31 & 3.07 &    1.58 &  1.54 &    1.01 &  1.17[12] &  0.77[9] &  0.25[4] &  0.53[5] & -0.53 \\
HD 87080   &         & -0.44 & 2.91 &    2.15 &  1.51 &    1.71 &  1.11[12] &  1.56[9] &  0.66[4] &  1.12[5] &  0.20 \\
HD 89948   &         & -0.30 & 2.68 &    1.43 &  1.14 &    0.85 &  1.02[12] &  0.65[9] &  0.16[4] &  0.43[5] & -0.29 \\
HD 92545   &         & -0.12 & 3.02 &    1.26 &  1.30 &    0.50 &  0.64[12] &  0.42[8] &  0.32[4] &  0.24[5] & -0.80 \\
HD 104979  & HR 4608 & -0.35 & 2.15 &    1.13 &  0.66 &    0.60 &  0.34[12] &  0.49[9] &  0.28[4] &  0.50[5] & -0.06 \\
HD 106191  &         & -0.29 & 3.25 & $<$1.09 &  1.70 & $<$0.50 &  0.91[12] &  0.48[6] &  0.20[3] &  0.46[5] & -1.20 \\
HD 107574  &         & -0.55 & 3.29 &    1.33 &  2.00 &    1.00 &  0.96[12] &  0.86[9] &  0.47[4] &  0.64[5] & -1.00 \\
HD 113226* & HR 4932 & +0.12 & 1.91 &    0.80 & -0.05 &   -0.20 & -0.08[12] & -0.10[9] &  0.02[4] &  0.01[5] & -0.15 \\
HD 116713  & HR 5058 & -0.12 & 2.53 &    1.76 &  0.81 &    1.00 &  0.96[12] &  0.80[9] &  0.40[4] &  1.06[5] &  0.19 \\
HD 116869  &         & -0.32 & 2.13 &    1.40 &  0.61 &    0.84 &  0.59[12] &  0.86[9] &  0.16[4] &  0.56[5] &  0.23 \\
HD 123396  &         & -1.19 & 1.56 &    1.50 &  0.91 &    1.81 &  0.70[12] &  1.55[9] &  0.50[4] &  1.19[5] &  0.90 \\
HD 139195  & HR 5802 & -0.02 & 2.07 &    1.11 &  0.25 &    0.25 &  0.39[12] &  0.08[9] &  0.18[4] &  0.10[5] &  0.00 \\
HD 147609  &         & -0.45 & 3.09 & $<$2.03 &  1.70 & $<$1.60 &  1.57[12] &  1.32[8] &  0.74[4] &  1.09[4] & -0.10 \\
HD 150862  &         & -0.10 & 2.99 &    1.48 &  1.25 &    0.70 &  1.08[12] &  0.34[8] &  0.20[3] &  0.23[4] & -0.55 \\
HD 181053  & HR 7321 & -0.19 & 1.85 &    0.97 &  0.20 &    0.27 &  0.34[12] &  0.20[9] &  0.15[4] &  0.15[5] &  0.05 \\
HD 188985  &         & -0.30 & 3.09 &    1.68 &  1.55 &    1.10 &  1.02[12] &  1.04[9] &  0.29[3] &  0.72[5] & -0.45 \\
HD 202109  & HR 8115 & -0.04 & 1.98 &    1.04 &  0.18 &    0.20 &  0.44[12] &  0.18[9] &  0.17[4] &  0.17[5] &  0.02 \\
HD 204075  & HR 8204 & -0.09 & 3.50 &    1.79 &  1.75 &    1.00 &  0.81[12] &  0.58[9] & -0.25[4] &  0.55[5] & -0.75 \\
HD 205011  &         & -0.14 & 2.10 &    1.32 &  0.40 &    0.58 &  0.71[12] &  0.38[9] &  0.21[4] &  0.36[5] &  0.18 \\
HD 210709  &         & -0.04 & 2.18 &    1.22 &  0.38 &    0.38 &  0.53[12] &  0.63[9] &  0.08[4] &  0.30[5] &  0.00 \\
HD 210910  &         & -0.37 & 1.47 & $<$0.69 &  0.00 & $<$0.18 &  0.54[9]  &  0.55[5] &  0.54[4] &  0.41[4] &  0.18 \\
HD 220009* & HR 8878 & -0.67 & 1.25 &    0.21 &  0.08 &    0.00 &  0.06[12] &  0.16[9] &  0.28[4] &  0.25[5] & -0.08 \\
HD 222349  &         & -0.63 & 2.89 &    1.55 &  1.68 &    1.30 &  1.03[12] &  1.26[9] &  0.24[2] &  0.88[5] & -0.38 \\
BD+18 5215 &         & -0.53 & 2.81 &    1.55 &  1.50 &    1.20 &  1.01[12] &  0.83[7] &  0.24[3] &  0.70[5] & -0.30 \\
HD 223938  &         & -0.35 & 2.13 &    1.58 &  0.64 &    1.05 &  0.74[12] &  1.08[9] &  0.37[4] &  0.75[4] &  0.41 \\
\noalign{\smallskip}
\hline
\end{tabular}
   $$
\end{table*}
%---------------------------------------------------------------------------------------

\section{The abundance results}\label{abund}

\begin{figure}[ht!]
\centerline{\includegraphics[totalheight=9.0cm]{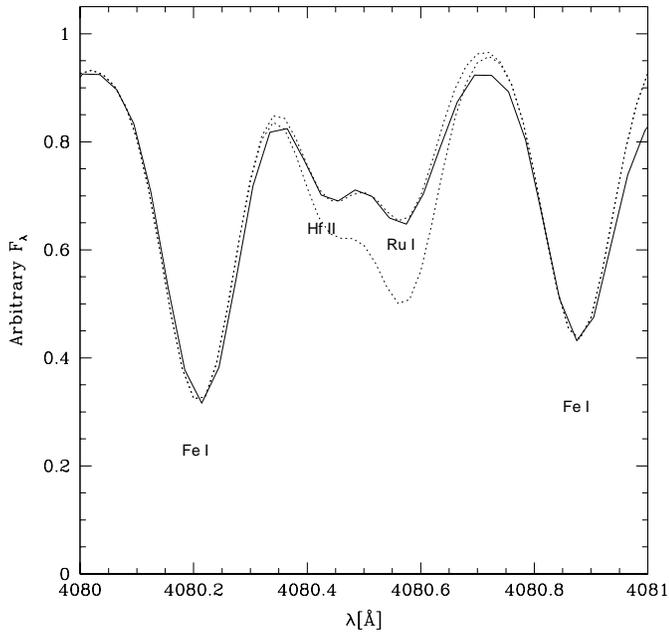}}
\caption{\label{6735fig4} Fitting of the \ion{Hf}{ii}
$\lambda$4080.437 and \ion{Ru}{i} $\lambda$4080.574 lines for the
barium star HD 181053. Solid line: observed spectrum; dotted lines: synthetic spectra with
[Hf/Fe] = 0.25, 0.40 and [Ru/Fe] = 0.15, 0.61.}
\end{figure}

\begin{figure}[ht!]
\centerline{\includegraphics[totalheight=9.0cm]{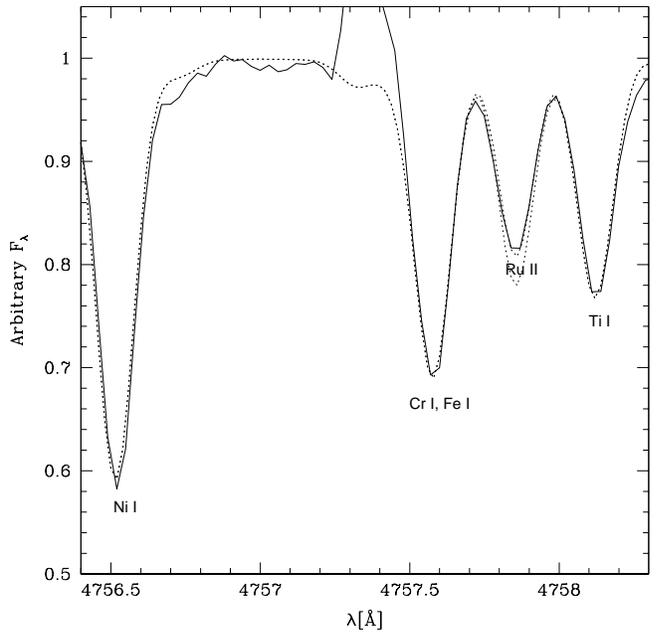}}
\caption{\label{6735fig5} Fitting of the \ion{Ru}{i} line
$\lambda$4757.856 for the barium star HD 87080.
Solid line: observed spectrum; dotted lines: synthetic spectra with
[Ru/Fe] = 1.60, 1.71.}
\end{figure}

The Ru and Hf transitions in the spectra of the barium stars of
our sample are generally weak, and this makes the abundance
determinations from them difficult. Figure \ref{6735fig4} shows
the synthetic spectrum fit in the star HD 181053 for the
$\lambda$4080.437 (\ion{Hf}{ii}), and $\lambda$4080.574
(\ion{Ru}{i}) lines. Figure \ref{6735fig5} shows the fit for the
$\lambda$4757.856 \ion{Ru}{i} line in the star HD 87080.

\begin{figure}[ht!]
\centerline{\includegraphics[totalheight=9.0cm]{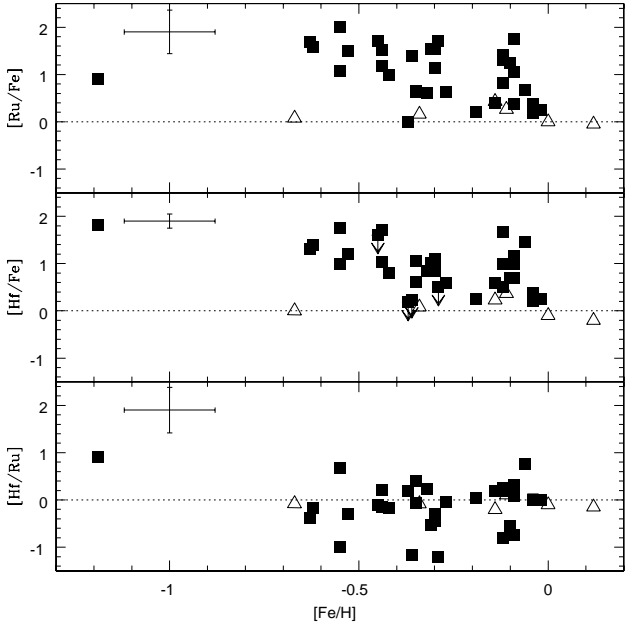}}
\caption{\label{6735fig6} [X/Fe] vs. [Fe/H] and [Hf/Ru] vs. [[Fe/H] for the sample stars.
Symbols: squares: barium stars; triangles: stars
considered normal rather than barium stars by \citet{rod07}. The arrows in the hafnium
panel indicate an upper limit for HR 107, HD 106191, HD 147609, and HD 210910. The
error bars indicate the maximum value of uncertainties on each axis.}
\end{figure}

The results for each line of our sample barium stars used for the 
abundance calculations are shown in Table \ref{eqwid}, and the 
average values are given in Table \ref{abuncom}. The latter shows that the 
average values are mainly in the ranges 
$+$0.18 $\leq$ [Ru/Fe] $\leq$ $+$2.00 and 
$+$0.20 $\leq$ [Hf/Fe] $\leq$ $+$1.71. Only for the star HD
210910 were the values lower than these ranges. Some stars
indicated in Tables \ref{eqwid} and \ref{abuncom} were considered 
normal rather than
barium stars by \citet{rod07} and their abundances were found to
be in the ranges 
$-$0.05 $\leq$ [Ru/Fe] $\leq$ $+$0.43 and 
$-$0.20 $\leq$ [Hf/Fe] $\leq$ $+$0.37. The upper values of these 
ranges in the
normal stars, when compared with those of the barium stars,
while showing the extent of the uncertainties, also highlight the
large overabundance of Ru and Hf in most barium stars of the
sample. In Fig. \ref{6735fig6} we plot the run of [Ru/Fe] and
[Hf/Fe] with [Fe/H] for all stars we analyzed. They show a
distinctive decreasing trend of both [Ru/Fe] and [Hf/Fe] towards
increasing [Fe/H] for the sample barium stars. It is noteworthy,
however, that the normal stars present no trend in
their [Ru/Fe] and [Hf/Fe] abundances, having [Ru,Hf/Fe] $\sim$ 0
over almost an order of magnitude variation in metallicity.

In some cases only an upper limit for the abundance could be derived, as
indicated in Tables \ref{eqwid} and \ref{abuncom}. If only one of the 
two lines has
an upper limit for the Ru abundance while the other has a good
fit, the adopted abundance was that resulting from the best line
and only this line was taken into account to compute the average
of the abundance, as seen comparing Tables \ref{eqwid} and \ref{abuncom}. 
For some stars the
fit at $\lambda$4080 is very uncertain and we chose not to derive
any abundance from it. The fit for BD+18 5215 at
$\lambda$4757 could not be carried out, and there is a spike of
noise at $\lambda$4093 for  HD 147609, so there are no results for
these transitions in these stars. For HD 20894, considered a
normal star by \citet{rod07}, and the barium stars HD 48565
and HD 76225, we found a large difference ($>$ 0.3 dex) between
the abundance results of the two Ru (barium stars) or Hf (HD
20894) lines. For all the other stars, the two lines used led to
similar abundance values for Ru as well as Hf. In Table \ref{comm}
the stars are ordered by increasing temperature, where 
the problems of the abundance derivation seem to be
related to higher temperatures. Although the temperatures of HD
210910 and HD 204075 are not very high, their broad lines cause
some difficulties in deriving the abundances. HD 106191 has an
upper limit for Hf, and no fit for the $\lambda$4080 line, yet its
temperature is not very high either. However, its S/N is lower
(S/N $\sim$ 100) than for other stars \citep[see][]{di06a,rod07}.
As a counter-example, the S/N $\sim$ 250 spectrum of HD 89948
allowed good fits for all lines, despite the high temperature of
this object. Regarding the lines of \ion{Ru}{i}, the fit of the
$\lambda$4757.856 line for the sample barium stars had higher
quality than for the $\lambda$4080.574 line due to its
freedom from neighboring perturbing lines.

%------------Table comm------------------------------------------------------------
\begin{table*}
\caption{This table was organized in increasing order of $T\rm _{eff}$, with the
comments about the results shown in the last column of each series of 4 columns.}
\label{comm}
   $$
\begin{tabular}{lcllclcllclcll}
\hline\hline
\noalign{\smallskip}
star & $T\rm _{eff}$ & $\log g$ & comm && star & $T\rm _{eff}$ & $\log g$ & comm  && star & $T\rm _{eff}$ & $\log g$ & comm\\
\noalign{\smallskip}
\hline
HD 123396  & 4360 & 1.4(3)  &         && HD 202109  & 4910 & 2.41    &      && HD 13551   & 5870 & 4.0(1)  &  \\
HD 220009  & 4370 & 1.91    &         && HD 104979  & 4920 & 2.58    &      && HD 106191  & 5890 & 4.2(1)  & a, c, g \\
HD 210910  & 4570 & 2.7(2)  & a, b, f && HD 46407   & 4940 & 2.65    &      && HD 8270    & 5940 & 4.2(1)  &  \\
HD 5424    & 4570 & 2.0(3)  &         && HD 223938  & 4970 & 3.1(1)  &      && HD 147609  & 5960 & 4.42(9) & a, b, e \\
HD 749     & 4610 & 2.8(1)  &         && HD 12392   & 5000 & 3.2(1)  &      && HD 89948   & 6010 & 4.30(8) & h \\
HD 210709  & 4630 & 2.4(2)  &         && HD 139195  & 5010 & 2.89    &      && HD 188985  & 6090 & 4.3(1)  & c \\
HD 26967   & 4650 & 2.51    &         && HD 20894   & 5080 & 2.60    & j    && HD 76225   & 6110 & 3.8(1)  & i \\
HD 116869  & 4720 & 2.2(2)  &         && HD 113226  & 5082 & 2.85    &      && HD 222349  & 6130 & 3.9(1)  &  \\
HD 205011  & 4780 & 2.41    &         && HD 13611   & 5120 & 2.49    &      && HD 92545   & 6210 & 4.0(1)  & c \\
HD 9362    & 4780 & 2.43    &         && HD 204075  & 5250 & 1.53    & c, f && BD+18 5215 & 6300 & 4.2(1)  & d \\
HD 116713  & 4790 & 2.67    &         && HD 22589   & 5400 & 3.3(1)  &      && HD 150862  & 6310 & 4.6(1)  & c \\
HD 181053  & 4810 & 2.48    &         && HD 87080   & 5460 & 3.7(2)  &      && HD 107574  & 6400 & 3.6(2)  & c \\
HD 27271   & 4830 & 2.9(1)  &         && HD 48565   & 5860 & 4.01(8) & i    && HR 107     & 6440 & 4.08(7) & a, b \\
\noalign{\smallskip}
\hline
\end{tabular}
   $$
Comments: a. upper limit for Hf; b. one of the lines has only an
upper limit for Ru; c. poor line profiles at $\lambda$4080; d. poor
line profiles of \ion{Ru}{i} at $\lambda$4757; e. poor line profile
of \ion{Hf}{ii} at $\lambda$4093; f. very broadened spectral
lines; g. S/N$\approx$100; h. S/N$\approx$250; i. $\Delta$(Ru) $>$
0.3; j. $\Delta$(Hf) $>$ 0.3.
\end{table*}
%---------------------------------------------------------------------------------------

From Fig. \ref{6735fig6}, the run of [Hf/Ru] with Fe
shows very large scatter, up to $\sim$2 dex, with an apparent advantage for
Ru abundances over ones of Hf, since most of the values of [Hf/Ru] lie below
zero. According to the Table \ref{abuncom}, among 33 barium stars, 17 have
[Hf/Ru] $<$ 0, 2 have [Hf/Ru] = 0, and 14 have [Hf/Ru] $>$ 0. Yet 
for normal stars of this table, 5 among 6 stars have [Hf/Ru] $<$ 0.
However, the highest values for [Ru/Fe] 
correspond mainly to stars that have only one line available and 
also when for one of the lines only an upper limit is available, 
which is much lower than the other estimate.
As shown in Sect. \ref{uncert}, uncertainties for [Ru/Fe] are 
much larger than ones for [Hf/Fe]. Furthermore, in cases where only one line 
was available, the error must be larger than for other stars 
when the two lines gave a result. The highest value of Ru abundance was
found for HD 107574. The spectral lines of this star are broader 
than those of most other stars, and, although the broadening is not
as strong as for HD 210910 and HD 204075, this line may be blended 
with another line that does not exist in our linelists, and it 
may be enough to give such high result. In fact, the line $\lambda$4757.856 
usually gives higher results than $\lambda$4080.574. Note that if the 
$gf$-value from VALD was used, the result would be higher. Hence, this
spectral region merits further spectroscopic analysis. A similar dificulty in 
determining abundances was found for the star HD 2454, with quite similar 
atmospheric parameters, in particular, microturbulent velocity.
According to the values of [X/Fe] for other neutron capture elements 
found for this star, these very low values of [Hf/Ru] may not be real.

\begin{figure*}[ht!]
\centerline{\includegraphics[totalheight=14cm]{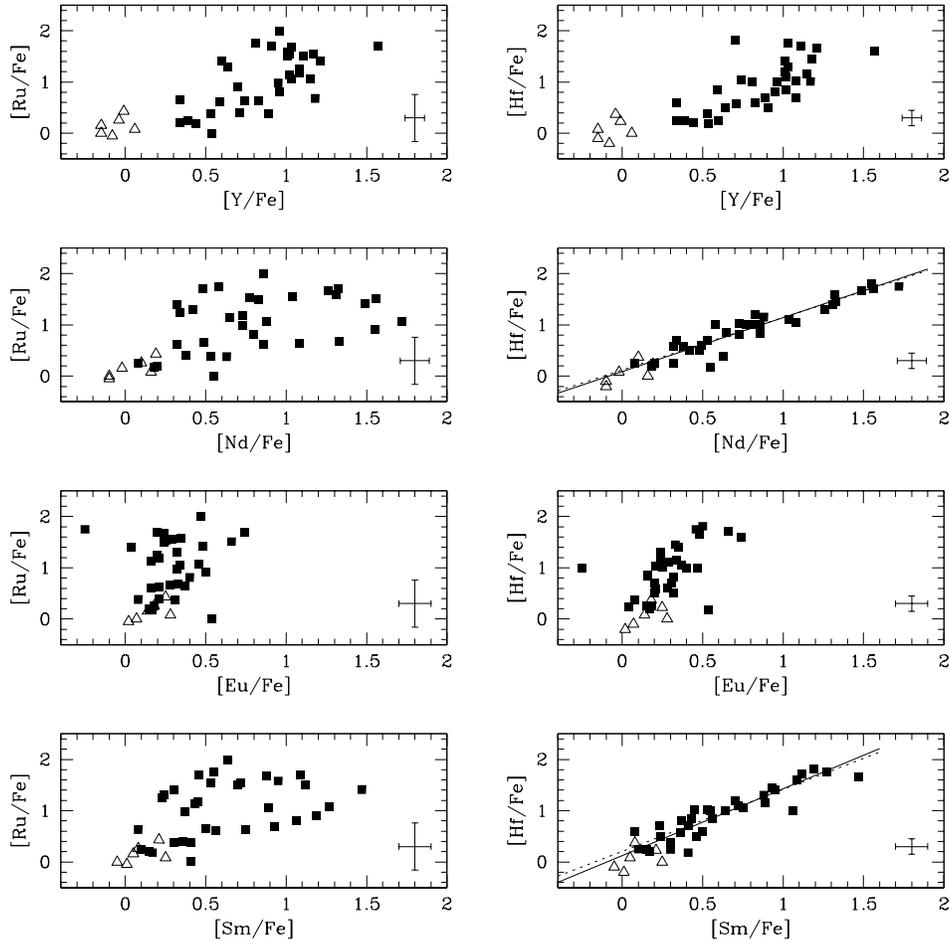}}
\caption{\label{6735fig7} Comparison of the [Hf/Fe] and [Ru/Fe]
behavior with [Y/Fe], [Nd/Fe], [Sm/Fe], and [Eu/Fe].
Filled squares are the barium stars and open triangles are those
considered normal rather than barium stars by \citet{rod07}. The least-square fits
for [Hf/Fe] vs. [Nd/Fe] and [Hf/Fe] vs. [Sm/Fe] are explained in the text.
The full line indicates the fit for all sample stars and the dashed line,
that excluding the normal stars.}
\end{figure*}

Figure \ref{6735fig7} shows the Ru and Hf abundances compared to
those of Y, Nd, Sm, and Eu for all stars of the
sample. Yttrium can properly represent the $s$-process, given
that, according to \citet{arlandini99}, 92\% and 8\% of its
abundance is due to $s$- and $r$-processes in the solar system,
respectively, as shown in Table \ref{frac}. In this table, the
missing abundance fractions are due to processes other than $s$
(main component) and $r$, and are seen to be of little significance.
Europium, compared to yttrium, has almost opposite behavior,
with 5.78\% and 94.25\% of its contribution from the $s$- and
$r$-processes, respectively. In their turn, neodymium and hafnium
may be considered as mild $s$-elements since the $s$-process forms the
bulk of their abundances but does not entirely dominate their
production, as it does for yttrium. Ruthenium and samarium,
on the other hand, are mainly contributed, but not entirely
dominated, by the $r$-process. These elements were chosen for a
comparison since they span a wide range of contributions from the
$s$- and $r$-processes, from a strong dominance of the $s$-process for
yttrium, through a more or less balanced contribution to
a high dominance of the $r$-process for europium.

To study the correlations between abundances shown in
Fig. \ref{6735fig7}, least-square fits are shown in Table
\ref{ajustes}. A remarkable feature of Fig. \ref{6735fig7} is 
the large scatter of Ru abundances, not found in the Hf
abundances. Also, Ru abundances seem to present little
correlation with those of other elements, as shown by the values
of $\chi^2_{red}$ in Table \ref{ajustes}.

The results for Hf have much lower scatter and are no less
remarkable: high scatter is found only in the [Hf/Fe] run with
[Y/Fe]. Rather tight positive correlations are found between the
Hf abundances and those of Nd and Sm, as can be judged by the
value of the $\chi^2_{red}$ of Table \ref{ajustes}. For Nd and
Sm, two fits for each are represented in Fig. \ref{6735fig7}: if all
sample stars are included in the fit, the results are:
\begin{equation}
\label{hfndcomp}
[{\rm Hf/Fe}] = (1.053 \pm 0.048)[{\rm Nd/Fe}] + (0.089 \pm 0.042)
\end{equation}
\begin{equation}
\label{hfsmcomp}
[{\rm Hf/Fe}] = (1.305 \pm 0.076)[{\rm Sm/Fe}] + (0.119 \pm 0.050).
\end{equation}

\noindent{If those objects considered normal rather than barium stars by \citet{rod07}
are withdrawn from the sample, the results are:}
\begin{equation}
\label{hfndbario}
[{\rm Hf/Fe}] = (1.022 \pm 0.055)[{\rm Nd/Fe}] + (0.121 \pm 0.050)
\end{equation}
\begin{equation}
\label{hfsmbario}
[{\rm Hf/Fe}] = (1.211 \pm 0.083)[{\rm Sm/Fe}] + (0.201 \pm 0.058).
\end{equation}

Nd and Sm however differ markedly in that Sm is fairly well
dominated by the $r$-process, whereas Nd shows a more balanced
fractional contribution in the solar system, similarly to Hf.
One should, on these grounds, expect a good correlation between
the [Hf/Fe] and [Nd/Fe] ratios, which we verify, and also a
reasonably clear correlation between the [Hf/Fe] and [Sm/Fe]
ratios, with higher slope, also verified by our data. The
correlation of the Hf abundances with Eu is distinctly less clear.
Since only $\sim$6\% of the solar abundance of Eu is from the
$s$-process, a very weak correlation between the [Hf/Fe] and [Eu/Fe]
ratios is to be expected, as we found, and only for higher values
of [Hf/Fe], above 1.5 dex or so, should a correlation be
detectable, but this lies outside the range of our data.

The main component of the $s$-process is believed to occur as a
chain from $\sp {56}$Fe seed nuclei up to Bi. The
neutron fluency may be enough to feed the first $s$-process
peak (near magic neutron number N = 50, in our discussion
represented by Y and Ru), then the second peak (near magic neutron
number N = 82, here Nd and Sm) and then on to the third
peak (near magic neutron number N = 126), as a function of the
so-called $^{13}$C pocket efficiency (still a free parameter in
current modelling), providing the bulk of the neutron flux (at
least for low fluxes) through the $^{\rm 13}$C($\alpha$,n)$^{\rm
16}$O reaction \citep{busso99}. It is thus possible that, in a
round of $s$-processing, not all peaks are equally fed by the
neutron fluency, generating scatter on the abundance ratios
involving elements from different peaks. Figure 16 of
\citet{busso99} illustrates this scatter on a diagram
of [hs/ls] vs. [Fe/H], as a function of the 
efficiency of the $^{13}$C pocket. For the [Ru/Fe] vs.
[Y/Fe] ratios we could in principle expect a good correlation,
since they both belong to the first peak, but Y is a magic neutron
element, for which an abundance enhancement is expected, 
partially masking this correlation. In fact, according to the
$\chi^2_{red}$ in Table \ref{ajustes}, the abundance correlation
between Ru and Y is only just as good as for Ru and Sm, and not
very clear. The [Ru/Fe] and [Eu/Fe] run demonstrates a very poor
correlation, not unexpected, these being elements from different
peaks, and with dissimilar fractional contributions from the 
$s$- and $r$-processes. The correlation between Ru (a first peak element)
and Nd or Sm is also expected to be worse under this reasoning,
since the latter are both near the second peak. Indeed, the run of
[Ru/Fe] and [Nd/Fe] does not show a good correlation, but that of
[Ru/Fe] and [Sm/Fe], somewhat unexpectedly, shows the hint of one,
as judged by the $\chi^2_{red}$ value. On the other hand, Hf, Nd,
and Sm are all near the second $s$-process peak, so one would expect
lower scatter in the abundance ratios shown in Fig. \ref{6735fig7},
also confirmed by our data. Some of the theoretical expectations
of AGB $s$-process nucleosynthesis are therefore borne out by our
results, but these, taken together, suggest a more complex
behavior of the abundance ratios of Ru and Hf with Y, Nd, Eu and
Sm than established by the current state of theory.
Clearly, more data on
the abundances of these two little-studied, spectroscopically not
very accessible, elements, are desirable to better
constrain theoretical scenarios of $s$-process nucleosynthesis.

%------------Table frac------------------------------------------------------------
\begin{table}
\caption{Contributions of $s$- and $r$-processes for the abundances of
Hf, Ru, Y, Eu, Sm, and Nd in solar system, following \citet{arlandini99}.
The last column is the sum of $s$- and $r$-processes.} \label{frac}
   $$
\begin{tabular}{rccc}
\hline\hline
\noalign{\smallskip}
el & $s$(\%) & $r$(\%) & $s$+$r$(\%) \\
\noalign{\smallskip}
\hline
Hf & 55.5  & 44.16 &  99.66 \\
Ru & 32.3  & 59.7  &  92 \\
Y  & 92    &  8    & 100 \\
Eu &  5.8  & 94.2  & 100 \\
Sm & 29.51 & 67.39 &  96.9 \\
Nd & 55.46 & 36.84 &  92.3 \\
\noalign{\smallskip}
\hline
\end{tabular}
   $$
\end{table}
%---------------------------------------------------------------------------------------

%------------Table ajustes------------------------------------------------------------
\begin{table}
\caption{Least-square fits, [X$_1$/Fe] = A[X$_2$/Fe] + B, where
X$_1$ is Ru or Hf and X$_2$ can be Y, Nd, Eu or Sm; 'cov' is
the covariance between A and B; 'd.o.f.' is the number of degrees of
freedom. Numbers in parenthesis are errors in last decimals.}
\label{ajustes}
   $$
\begin{tabular}{rrcllclc}
\hline\hline
\noalign{\smallskip}
X$_1$ & X$_2$ && A & B & $\chi^2_{red}$ & cov & d.o.f. \\
\noalign{\smallskip}
\hline
\multicolumn{3}{c}{} & \multicolumn{5}{c}{all stars} \\
\noalign{\smallskip}
\hline
Ru &  Y && 0.778(103) & 0.630(103) & 3.4 & -0.010 & 37 \\
Ru & Nd && 0.443(70)  & 1.024(64)  & 4.0 & -0.004 & 37 \\
Ru & Eu && 0.707(173) & 1.152(61)  & 4.1 & -0.009 & 37 \\
Ru & Sm && 0.656(100) & 0.981(67)  & 3.4 & -0.006 & 37 \\
Hf &  Y && 1.036(53)  & 0.005(49)  & 6.1 & -0.002 & 37 \\
Hf & Nd && 1.053(48)  & 0.089(42)  & 1.3 & -0.002 & 37 \\
Hf & Eu && 1.778(188) & 0.333(62)  & 4.2 & -0.010 & 37 \\
Hf & Sm && 1.305(76)  & 0.119(50)  & 1.5 & -0.003 & 37 \\
\noalign{\smallskip}
\hline
\multicolumn{3}{c}{} & \multicolumn{5}{c}{without normal stars} \\
\noalign{\smallskip}
\hline
Ru &  Y && 0.571(123) & 0.848(125) & 3.9 & -0.015 & 31 \\
Ru & Nd && 0.333(73)  & 1.137(67)  & 4.0 & -0.004 & 31 \\
Ru & Eu && 0.559(168) & 1.234(60)  & 3.8 & -0.009 & 31 \\
Ru & Sm && 0.516(100) & 1.095(68)  & 3.5 & -0.006 & 31 \\
Hf &  Y && 1.216(82)  & -0.174(80) & 6.4 & -0.006 & 31 \\
Hf & Nd && 1.022(55)  & 0.121(50)  & 1.4 & -0.002 & 31 \\
Hf & Eu && 1.474(175) & 0.503(61)  & 4.5 & -0.009 & 31 \\
Hf & Sm && 1.211(83)  & 0.201(58)  & 1.5 & -0.004 & 31 \\
\noalign{\smallskip}
\hline
\end{tabular}
   $$
\end{table}
%---------------------------------------------------------------------------------------

%
%______________________________________________________________

\section{Conclusions}\label{concl}

We present abundances of Ru and Hf determined through the spectrum
synthesis of two lines for each element available in the
spectra of dwarf and giant barium stars. Good agreement was
obtained for each pair of Ru and Hf lines for most
sample stars. For a few stars, the abundance
difference derived from the two lines was higher than 0.3
dex. We found that for the $\lambda$4080.437 of
\ion{Hf}{ii}, and $\lambda$4757.856 of \ion{Ru}{i} lines,
published $gf$-values do not fit well to the observed solar spectrum.
We offer tentative explanations for these discrepancies,
which merit further study. New experiments determining the $\log gf$ 
for these lines are needed. Further, reliable abundance
determinations of these elements can contribute 
considerably to our knowledge of heavy element abundances
in this class of chemically peculiar stars, besides helping
better constrain theoretical scenarios of AGB $s$-process
nucleosynthesis, still prone to important uncertainties.

We compared the run of [Hf/Fe] and [Ru/Fe] abundance ratios
with other heavy elements, chosen to represent
different fractional contributions from the $s$- and $r$-process, as
judged by the solar system isotopic composition. The abundance of
Hf is closely correlated with that of Sm and Nd, in
reasonable agreement with theoretical expectations. It is
noteworthy that, although Sm is an $r$-process dominated element,
and Nd presents abundance fractions from $s$- and $r$-process
nucleosynthesis similar to Hf in the solar system abundance
pattern, in our barium star data both elements are well 
correlated with Hf, probably because all three elements
lie near the second $s$-process peak. Ru is not
clearly correlated with the other heavy elements, excepting
possibly Y and Sm. A correlation of Ru and Y abundances may be
masked partially by the magic neutron number nature of the latter.
The possibility of a correlation in the Ru and Sm abundances, in
the light of a similar fractional contribution from the 
$s$- and $r$-processes even though these elements belong to different
$s$-processing peaks, deserves further investigation. The stars should
span a larger metallicity interval than the one studied 
here. These results suggest a more complex relationship between
the excesses of the various heavy elements in barium stars 
than implied by theoretical considerations.

%
%______________________________________________________________

\begin{acknowledgements}
DMA acknowledges a FAPERJ post-doctoral fellowship n$^{\circ}$
152.680/2004, as well as CAPES, for the post-doctoral fellowship
n$^{\circ}$ BEX 3448/06-1.
We are also grateful to Licio da Silva, Luciana
Pomp\'eia, Paula Coelho, and Jorge Mel\'endez for carrying out
some observations of our sample spectra. We are grateful to
Beatriz Barbuy for make available part of the spectra and the
spectrum synthesis code. GFPM acknowledges financial support by
CNPq/Conte\'udos Digitais (grant 552331/01-5),
CNPq/MEGALIT/Institutos do Mil\^enio program, and a FAPERJ (grant
APQ1/26/170.687/2004). We thank the referee,
Dr. Roberto Gallino, for his criticism and comments, which very
considerably improved this paper.
\end{acknowledgements}
%
%______________________________________________________________


\begin{thebibliography}{}

\bibitem[Allen \& Barbuy(2006a)]{di06a}Allen, D.M., Barbuy, B. 2006, A\&A, 454, 895

\bibitem[Allen \& Barbuy(2006b)]{di06b}Allen, D.M., Barbuy, B. 2006, A\&A, 454, 917

\bibitem[Arlandini  et al.(1999)]{arlandini99}Arlandini, C., K\"appeler, F., Wisshak, K. 1999, ApJ, 525, 886

\bibitem[Barbuy et al.(2003)]{barb03}Barbuy, B., Perrin, M.-N., Katz, D., Coelho, P., Cayrel, R., Spite, M.,
   Van't Veer-Menneret, C. 2003, A\&A, 404, 661

\bibitem[B\"ohm-Vitense et al.(1984)]{BV84} B\"ohm-Vitense, E., Nemec, J., Proffit, C. 1984, ApJ, 278, 726

\bibitem[B\"ohm-Vitense et al.(2000)]{bohm2000}B\"ohm-Vitense, E., Carpenter, K., Robinson, R., Ake, T.,
   Brown, J. 2000, ApJ, 533, 969

\bibitem[Boyarchuk et al.(2002)]{boyarchuk02}Boyarchuk, A.A., Pakhomov, Yu. V., Antipova, L.I.,
   Boyarchuk, M.E. 2002, ARep, 46, 819

\bibitem[Busso et al.(1999)]{busso99}Busso M., Gallino, R., Wasserburg G.J. 1999, ARA\&A, 37, 239

\bibitem[Cayrel et al.(1991)]{cay91}Cayrel, R., Perrin, M.N., Barbuy, B., Buser, R. 1991, A\&A, 247, 108

\bibitem[Edvardsson et al.(1993)]{edv93}Edvardsson, B., Andersen, J., Gustafsson, B., Lambert, D.L.,
   Nissen, P.E., Tomkin, J. 1993, A\&A, 275, 101

\bibitem[Grevesse \& Sauval(1998)]{gs98}Grevesse, N., Sauval, A.J. 1998, Space Sci. Rev., 85, 161

\bibitem[Gustafsson et al.(1975)]{gben75}Gustafsson B., Bell K.A., Eriksson K., Nordlund \AA. 1975, A\&A, 42, 407

\bibitem[Hill et al.(2002)]{hill02}Hill, V., Plez, B., Cayrel, R., et al. 2002, A\&A, 387, 560

\bibitem[Hinkle et al.(2000)]{hinkle00}Hinkle, K., Wallace, L., Valenti, J., Harmer, D. 2000,
   Visible and Near Infrared Atlas of the Arcturus Spectrum 3727-9300 \AA

\bibitem[Honda et al.(2006)]{honda06}Honda, S., Aoki, W., Ishimaru, Y., Wanajo, S., Ryan, S.G. 2006,
   ApJ, 643, 1180

\bibitem[Kaufer et al.(2000)]{kaufer00}Kaufer, A., Stahl, O., Tubbesing, S., et al. 2000, Proc. SPIE, 1008, 459

\bibitem[Kurucz et al.(1984)]{kurucz84}Kurucz, R. L., Furenlid, I., Brault, J. 1984, Solar flux atlas from 296
   to 1300 nm, National Solar Observatory Atlas, Sunspot (New Mexico: National Solar Observatory)

\bibitem[Jorissen et al.(1998)]{J98}Jorissen, A., Van Eck, S., Mayor, M., Udry, S. 1998, A\&A, 332, 877

\bibitem[Lawler et al.(2007)]{lawler07}Lawler, J.E., Den Hartog, E.A., Labby, Z.E., Sneden, C.,
   Cowan, J.J., Ivans, I.I. 2007, ApJS, 169, 120

\bibitem[Luck \& Bond(1991)]{LB91}Luck, R.E., Bond, H.E. 1991, ApJS, 77, 515.

\bibitem[Lundqvist et al.(2006)]{lund06}Lundqvist, M., Nilsson, H., Wahlgren G.M., Lundberg, H., Xu, H.L.,
   Jang, Z.-K., Leckrone, D.S. 2006, A\&A, 450, 407

\bibitem[Masseron et al.(2006)]{mass06}Masseron, T., Van Eck, S., Famaey, B., Goriely, S.,
   Plez, B., Siess, L., Beers, T.C., Primas, F., Jorissen, A. 2006, A\&A, 455, 1059

\bibitem[Mel\'endez et al.(2003)]{mel03}Mel\'endez, J., Barbuy, B., Bica, E., Zoccali, M., Ortolani, S.,
   Renzini, A., Hill, V. 2003, A\&A, 411, 417

\bibitem[North et al.(1994)]{north94} North, P., Berthet, S., Lanz, T. 1994, A\&A, 281, 775

\bibitem[Piskunov et al.(1995)]{pisk95}Piskunov, N., Kupka, F., Ryabchikova, T., Weiss, W.,
   Jeffery, C. 1995, A\&AS, 112, 525

\bibitem[Plez et al.(1992)]{plez92}Plez, B., Brett, J.M., Nordlund, A. 1992, A\&A, 256, 551

\bibitem[Smiljanic et al.(2007)]{rod07}Smiljanic, R., Porto de Mello, G.F., da Silva, L. 2007, A\&A, 468, 679

\bibitem[Smith et al.(1996)]{scjb96}Smith, V.V., Cunha, K., Jorissen, A., Boffin, H.M.J. 1996, A\&A, 315, 179

\bibitem[Sneden et al.(2003)]{sneden03}Sneden C., Cowan J.J., Lawler J.E., et al. 2003, ApJ, 591, 936

\bibitem[Spite(1967)]{Spi67}Spite, M. 1967, Ann.. Astrophys. 30, 211

\bibitem[Tomkin \& Lambert(1983)]{tl83}Tomkin J., Lambert D.L. 1983, ApJ, 273, 722

\bibitem[Yushchenko et al.(2002)]{yush02}Yushchenko, A., Gopka, V., Kim, C., Khokhlova, V., et al.
   2002, JKAS, 35, 209

\end{thebibliography}
\end{document}